\documentclass{article}

\usepackage[english]{babel}

\usepackage[letterpaper,top=2cm,bottom=2cm,left=3cm,right=3cm,marginparwidth=1.75cm]{geometry}

\usepackage{amsmath}
\usepackage{graphicx}
\usepackage[colorlinks=true, allcolors=blue]{hyperref}
\usepackage{epsfig}
\usepackage{subcaption}
\usepackage{calc}
\usepackage{amssymb}
\usepackage{amstext}
\usepackage{amsmath}
\usepackage{amsthm}
\usepackage{multicol}
\usepackage{pslatex}
\usepackage{apalike}
\usepackage{algorithm2e}
\usepackage[bottom]{footmisc}
\usepackage{cite}
\usepackage{hyperref}
\usepackage{framed,multirow}
\usepackage{comment}
\usepackage{graphicx}
\usepackage{subcaption}
\usepackage{lipsum}
\usepackage{esint}
\usepackage{mwe}
\usepackage{mathtools}
\usepackage{caption}
\usepackage{cuted}
\usepackage{capt-of}
\usepackage{booktabs}
\usepackage{makecell} 
\usepackage[export]{adjustbox}
\usepackage{subcaption}

\title{Data-driven Viscosity Solver for Fluid Simulation}
\author{Wonjung Park, Hyunsoo Kim, Jinah Park\\
\normalsize{Korea Advanced Institute of Science and Technology, KAIST}\\\normalsize{fabiola@kaist.ac.kr}}
\date{}
\begin{document}
\maketitle

\begin{abstract}
We propose a data-driven viscosity solver based on U-shaped convolutional neural network to predict velocity changes due to viscosity. Our solver takes velocity derivatives, fluid volume, and solid indicator quantities as input.
The traditional marker-and-cell (MAC) grid stores velocities at the edges of the grid, causing the dimensions of the velocity field vary from axis to axis.
In our work, we suggest a symmetric MAC grid that maintains consistent dimensions across axes without interpolation or symmetry breaking.
The proposed grid effectively transfers spatial fluid quantities such as partial derivatives of velocity, enabling networks to generate accurate predictions.
Additionally, we introduce a physics-based loss inspired by the variational formulation of viscosity to enhance the network's generalization for a wide range of viscosity coefficients.
We demonstrate various fluid simulation results, including 2D and 3D fluid-rigid body scenes and a scene exhibiting the buckling effect. Our code is available at \url{https://github.com/SSTDV-Project/python-fluid-simulation.}
\end{abstract}

\section{\uppercase{Introduction}}
\label{sec:introduction}

Numerical methods for fluid simulation have been widely used in various fields such as industrial design, and computer graphics.
These methods solve incompressible Navier-Stokes (N-S) equations, the governing equation of fluid dynamics in the form of partial differential equations (PDEs).
The process of solving N-S equations is well-known for being computationally expensive and the cost is mostly from solving a series of linear systems by iterative methods.

% ML + fluid simulation (projection) - why projection, but not viscosity?
In recent years, various studies \cite{kim2019deep, richter2022neural, Accelerating} have adopted machine learning to accelerate the projection step of fluid simulation.
By replacing iterative methods with highly parallel GPU-optimized neural networks, they were able to greatly accelerate the overall fluid simulation.
In addition to the parallelism, these methods exploit the fact that pressure makes the fluid incompressible.
Unlike typical machine learning regression, which solely relies on supervised learning from the dataset, they are designed to enforce the incompressibility of the velocity field.
Therefore, they can be trained with a small dataset and generalize well to scenes outside of the training data.

% Problem Statement: Why no viscosity solver yet?
Although the data-driven projection solver accelerates the simulation of inviscid fluid to the real-time level, we found that it is insignificant when used for a viscous fluid.
Traditional viscosity solvers have to solve linear systems that are 4 times larger than the projection stage in 2D (x8 for 3D), since it deals with twice more fluid quantities per each dimension.
However, studies on the data-driven viscosity solver have not yet been conducted in depth, since generalization of the viscosity solver is a challenging problem.

The challenge comes from two factors.
First, it is difficult for networks to generalize over viscosity coefficients because it varies in a wide range from zero (inviscid) to infinity (solid).
Second, the pressure is stored at the cell center whereas the fluid velocity is stored at the cell boundary of the marker-and-cell (MAC) grid for Eulerian simulation. This structure yields an asymmetry of the velocity field; for example, the velocity of the x-direction has the dimension $(n+1, n)$ but the y-direction has $(n, n+1)$ for the Eulerian MAC grid $n \times n$. 
Therefore, unlike general image data for CNNs, fluid velocity requires special care due to their distinct and asymmetric structure.

We design the viscosity solver architecture with these problems in mind.
To generalize the solver for viscosity coefficients, we inherit the variational interpretation of the viscosity from the conventional solver \cite{Batty}, using its minimization problem as the training loss.
In addition, we introduce the \emph{symmetric MAC grid} to enable the symmetric representation of fluids regardless of the axis.

To our knowledge, this is the first data-driven viscosity solver, so we focus on elaborating an architecture that generalizes to various scenarios.
In this paper, we validate our architecture through ablation studies of neural network configurations.
Finally, we demonstrate the universality of our solver in various scenarios, including unseen rigid-fluid scenes, mixed dynamic viscosity scenes, and 3D scene which shows buckling effect of viscous fluid.

\section{Related Work}

The main subject of this study is to simulate viscous fluids governed by the incompressible Navier-Stokes (N-S) equations with our data-driven viscosity solver.
The Navier-Stokes (N-S) equations are:
{\small
\begin{equation}
    \frac{D\textbf{v}}{Dt} = -\frac{1}{\rho} \nabla p + \frac{\mu}{\rho} \nabla^2 \textbf{v} + \textbf{g}, \label{eq:n-s-momentum}
\end{equation}
\begin{equation}
    \nabla \cdot \textbf{v} = 0. \label{eq:n-s-incompressible}
\end{equation}
}
There are three approaches to numerically simulate the N-S equations: grid-based (Eulerian), particle-based (Lagrangian), and Lagrangian/Eulerian hybrid appraoches.
Hybrid simulation uses particles to transport fluid and grids to update velocity by external forces, pressure, and viscosity.
For spatial discretization of fluid quantities, marker-and-cell method \cite{MAC}, which stores velocities on edges of cells, is often used in grid-based simulation. 

The projection term in the N-S equations (the first term in the RHS of \autoref{eq:n-s-momentum}) enforces the incompressibility as formulated in \autoref{eq:n-s-incompressible}.
Combining the two equations gives Poisson's equation ($\nabla^2 p = f$), whose discrete form can be solved using linear system solvers.

The viscosity term in the N-S equations (second term in RHS of \autoref{eq:n-s-momentum}) gives the viscous behavior, accounting for internal friction.
With explicit solvers, this term can become numerically unstable especially for high dynamic viscosity $\mu$.
Thus, a more stable implicit formulation of the viscosity fluid for hybrid methods is proposed in \cite{Batty}. They interpreted the viscosity term as a minimization problem that seeks to minimize the rate of viscous dissipation with the smallest velocity change possible. In 2D, the minimization problem is defined as:
{\small\begin{equation}\label{equ:min_prob}
\min_{\overrightarrow{u}}\iint \!\rho \left \| \overrightarrow{u}\!-\!\overrightarrow{u}^{old} \right \|^{2}\!+\!2\Delta t\iint\!\mu\left\|\frac{\nabla\overrightarrow{u}+(\nabla\overrightarrow{u})^{T})}{2} \right \|_{F}^{2}
\end{equation}}

To reduce the computational cost of fluid simulations, various dimensionality reduction methods have been proposed. Model reduction for fluid-rigid simulation is discussed in \cite{ModelReduction}.
In addition, multi-resolution Eulerian grids are introduced for efficient Poisson equation solving in both inviscid \cite{TallCell, Multigrid} and viscous fluids \cite{aanjaneya2019efficient, shao2022fast}.
Furthermore, recent studies applied machine learning techniques including principal component analysis \cite{xiao2018novel}, regression forests \cite{Regression}, Long Short-Term Memory \cite{latent}, generative adversarial networks \cite{DCGAN}, and graph networks \cite{sanchez2020learning, li2022graph}.

Among these machine learning techniques, convolutional neural networks (CNNs) have shown their effectiveness in Eulerian and hybrid fluid simulation.
Since the Eulerian grid has a similar structure to the image, many studies applied CNN to fluid simulation \cite{CNNsteady, CFDNet, RANS, Correction}.
Some studies \cite{kim2019deep} proposed generative networks that output the next velocity fields based on reduced parameters.
Alternatively, other studies \cite{Accelerating, xiao2018novel, Adaptive, Projection, gao2020accelerating} maintain the traditional fluid simulation process but replace the most time-consuming projection step with a data-driven solver.
The traditional projection step with discretized PDE is computationally expensive and data dependent, since it is solved by iterative methods until it meets the conditions.
Thus, the replacement of Poisson's equation with deep learning techniques \cite{richter2022neural, deepPoisson} have been actively researched to efficiently derive the solution.
Despite the efficiency and generalization of these data-driven projection solvers, their impact on viscous fluids is limited as the viscosity step remains a significant bottleneck.

In this study, we propose a CNN-based viscosity solver that replaces the PDE-based solver \cite{Batty} to effectively reduce the computational costs of the viscous fluid.
%Our simulator is based on affine particle-in-cell (APIC) \cite{APIC} with a CNN-based data-driven viscosity solver.

\section{Method}

\subsection{Data-driven viscosity solver}
\begin{figure}[!h]
\centering
\includegraphics[width=0.6 \textwidth]{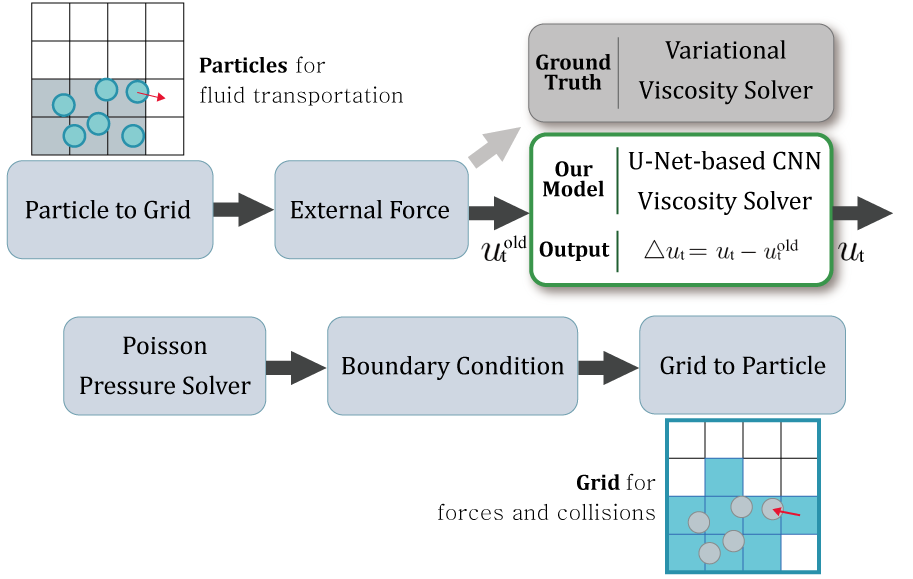}
\caption{\textbf{Fluid simulation with data-driven viscosity solver.} Using the traditional affine particle-in-cell \cite{APIC} simulator as a baseline, our data-driven solver replaces the viscosity solver to output velocity changes due to viscosity. Training ground truth is generated with the traditional viscosity solver \cite{Batty}.}
\label{fig:overall}
\end{figure}

\begin{figure}[h]
\centering
\includegraphics[width=0.5\textwidth]{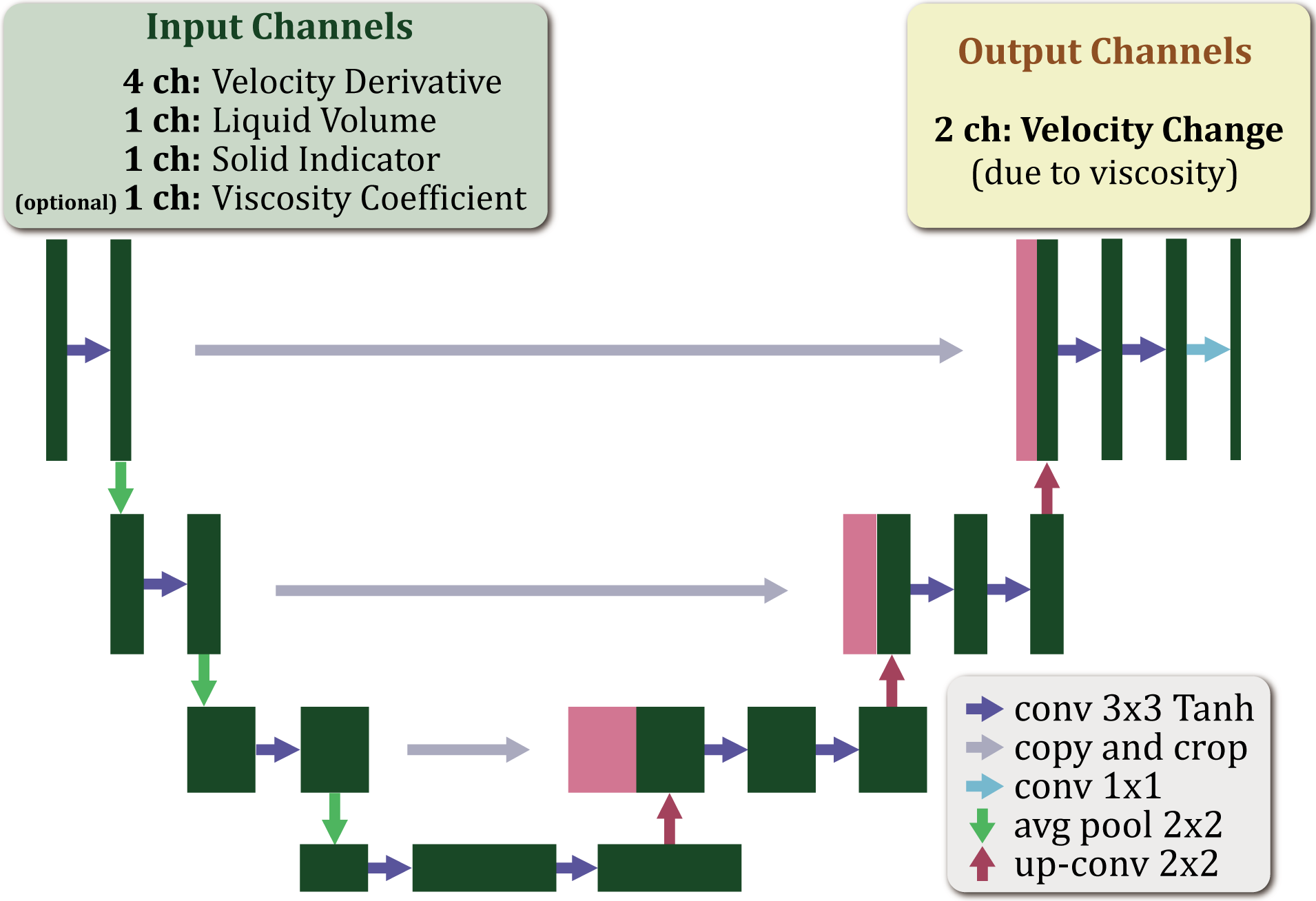}
\caption{U-shaped convolutional neural network for our data-driven viscosity solver.}
\label{fig:network}
\end{figure}

We suggest a data-driven viscosity solver that predicts velocity change by viscosity.
As \autoref{fig:overall} shows, the overall process is based on the  affine particle-in-cell (APIC) \cite{APIC} simulator, which transports fluid with particles and updates the velocity by external force, viscosity, and pressure on the grid.
The viscosity solver updates the velocity $u_{t}^{old}$ from the external force step (gravity) to the velocity $u_{t}$.
We use the velocity difference $\triangle u_{t}=u_{t}-u_{t}^{old}$ solved by the traditional solver \cite{Batty} as the ground truth for training.

\subsubsection{Loss function}
We train our model by supervised learning with L2 loss which is the mean squared error between ground truth and model output.
In addition, we suggest the \emph{variational loss} ${L}_{v}$ which is the variational formulation of the conventional viscosity solver.
When training the model only for fixed dynamic viscosity, L2 loss is sufficient to converge the model.
However, for a dataset that contains various dynamic viscosity at the same time, we use both losses together to regularize the network for dynamic viscosity $\mu$. Our loss function is composed of the sum of L2 loss and variational loss ${L}_{v}$ with equal weights.
We trained the network with Adam optimizer\cite{kingma2014adam}. The variational loss ${L}_{v}$ on the 2D Eulerian grid is derived as follows, where velocity $\textbf{u}=(u, v)$:
\begin{equation}\label{equ:dissipation}
\resizebox{.8\hsize}{!}{$
\begin{split}
&{L}_{v}(\textbf{u},\triangle t, \rho, \mu) \\
&=\frac{1}{n(n+1)}  \sum_{i=1}^{n+1}\!\sum_{j=1}^{n}\rho
 \Bigl(u_{i\!-\!\frac{1}{2}, j}\!-\!u^{old}_{\ i\!-\!\frac{1}{2},j} \Bigl)^{2}
+\frac{1}{n(n+1)}  \sum_{i=1}^{n}\!\sum_{j=1}^{n+1}\rho
 \Bigl(v_{i, j\!-\frac{1}{2}} \!-\! v^{old}_{\ i,j\!-\!\frac{1}{2}} \Bigl)^{2}+\frac{1}{n^{2}}2\triangle t\sum_{i=1}^{n}\!\sum_{j=1}^{n}\mu \left \| \frac{(\triangledown {u}_{i,j} + \triangledown {u}_{i,j}^{T})}{2}\right \|^{2}_{F}\\
&\textrm{where} \ \triangledown {u}_{i,j}\!=\!\begin{bmatrix}
\frac{1}{\triangle x}({u}_{i+\frac{1}{2},j}\!-\!{u}_{i-\frac{1}{2},j})
& \frac{1}{\triangle y}({u}_{i,j+\frac{1}{2}}\!-\!{u}_{i,j-\frac{1}{2}})\\
\frac{1}{\triangle x}({v}_{i+\frac{1}{2},j}\!-\!{v}_{i-\frac{1}{2},j})
& \frac{1}{\triangle y}({v}_{i,j+\frac{1}{2}}\!-\!{v}_{i,j-\frac{1}{2}})
\end{bmatrix}
\end{split}
$}
\end{equation}

The sum of the first and second term of ${L}_{v}$ can be derived by the average of the element-wise squared output because the output is velocity change due to viscosity. The third term denotes the fluid dissipation over the timestep. Through this term, ${L}_{v}$ imposes a greater loss for higher viscosity, helping the network to create less velocity difference between adjacent fluid elements.

\begin{figure*}[t]
\captionsetup{font={footnotesize}}

\centering
\begin{subfigure}{0.32\textwidth}
    \centering
    \includegraphics[width=\textwidth]{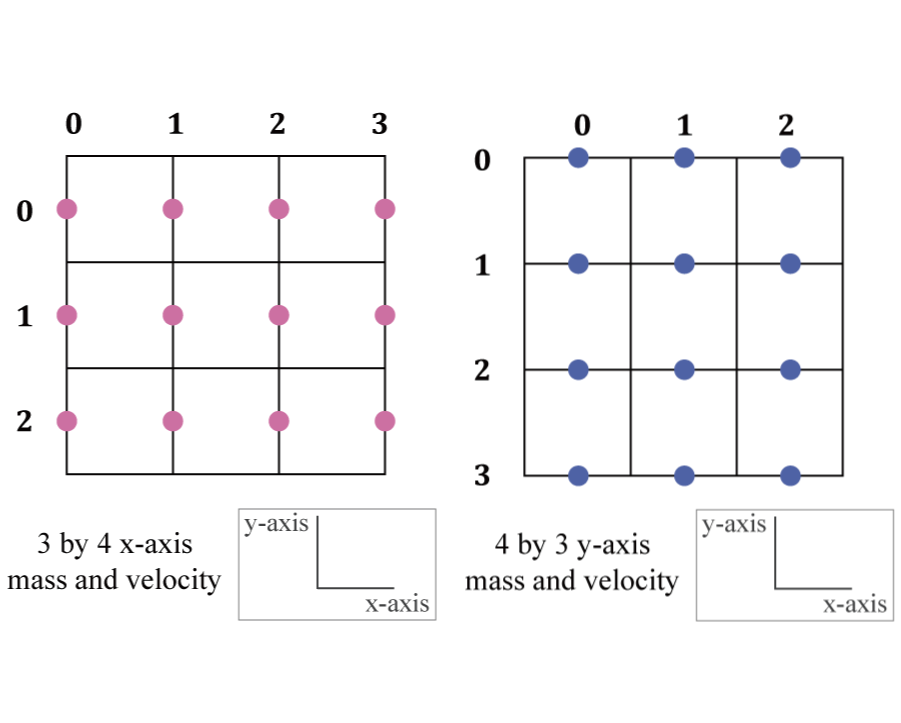}
    \caption{Marker-and-cell (MAC) grid}
    \label{fig:sym-a}
\end{subfigure}
\begin{subfigure}{0.42\textwidth}
    \centering
    \includegraphics[width=\textwidth]{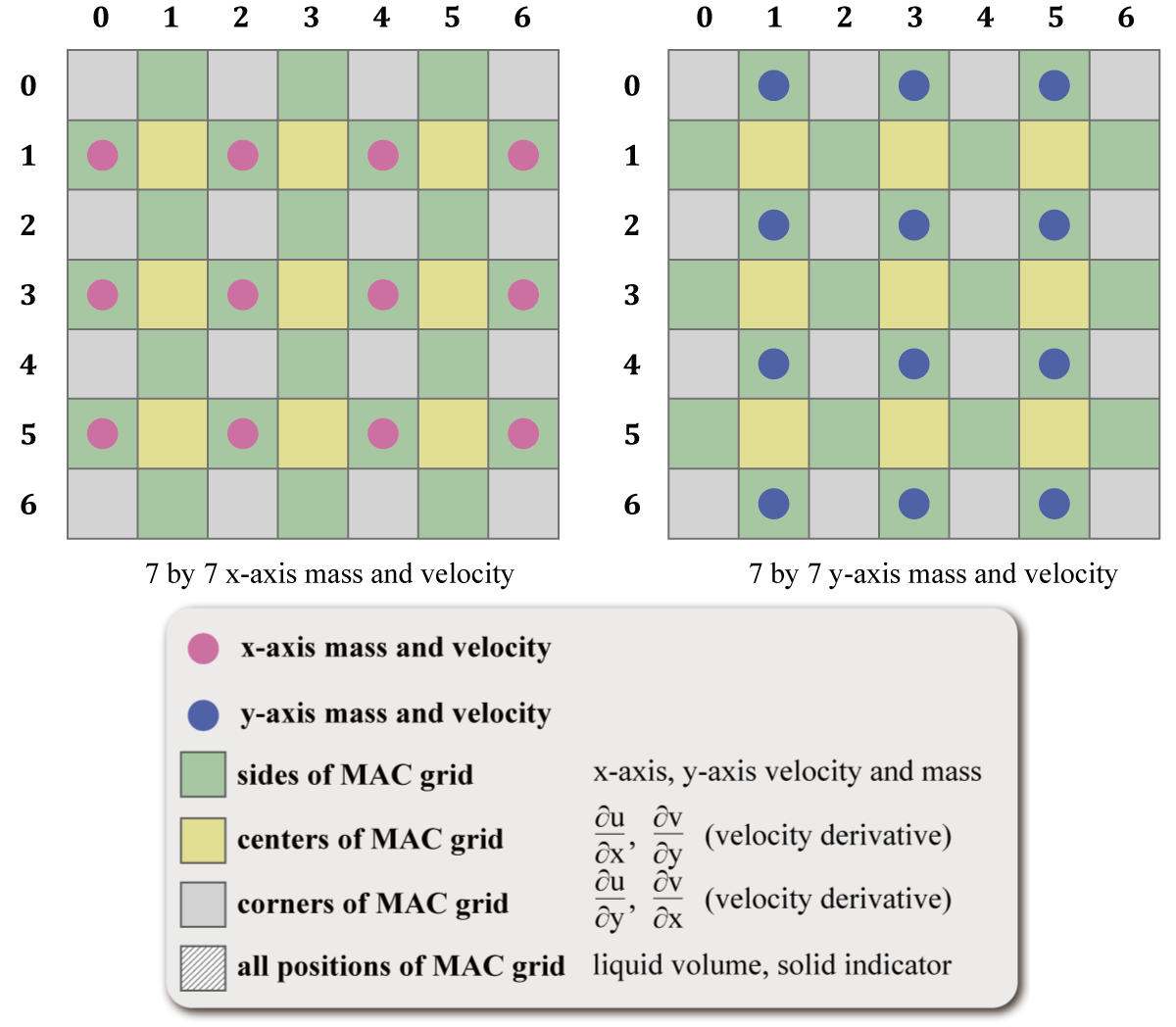}
    \caption{Symmetric MAC grid for 2D CNN}
    \label{fig:sym-b}
\end{subfigure}
\begin{subfigure}{0.24\textwidth}
    \centering
    \includegraphics[width=0.62\textwidth]{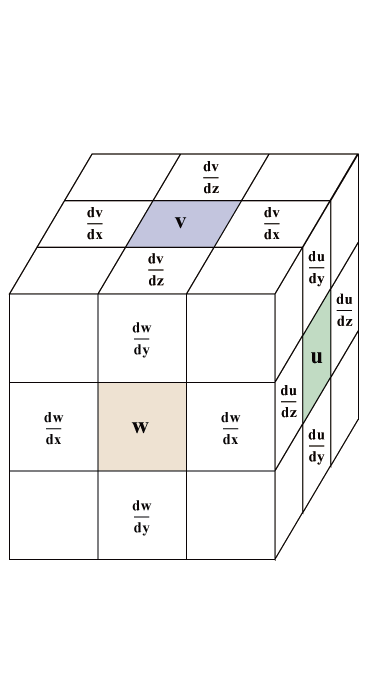}
    \caption{Symmetric MAC grid for 3D CNN}
    \label{fig:3d}
\end{subfigure}
\caption{\textbf{Extension of the MAC grid for CNN-based viscosity solver} To adjust input channel dimensions without losing information or breaking symmetry, we propose the symmetric MAC grid. This grid ensures symmetry in all input channels.}
\label{fig:figures}
\end{figure*}

\subsection{Architecture}

\subsubsection{U-shaped CNN model}
We design the neural network architecture with the following two conditions in mind.
First, the network allows input and output to have the same dimensions in width and height since they are defined on the same Eulerian grid.
Second, the architecture should grasp important features of viscosity.

Viscosity is the resistance force of a fluid that reduces the velocity difference with neighboring fluids.
In addition, fluids in the boundary layer, which is a thin layer near the boundary between the fluid and the solid, are more affected by viscosity.
In other words, both the local feature representing adjacent fluid elements and the global feature revealing the location of the fluid element with respect to the fluid-rigid body scene are important for predicting the velocity.
Therefore, we set the U-Net \cite{UNet} as a backbone because its multi-resolution feature maps help the network learn both local and global features of fluid-rigid body scene, and generate output of the same size as the input.

Unlike the image segmentation task of the original U-Net, our viscosity solver aims to output the velocity change for each dimension.
The velocity change has the distinct property that it has the range $(-\infty, \infty)$.
Constructing a network that produces exact velocity changes that could be positive or negative is a challenging task.
To find the best architecture, we have experimented with various conditions to explore which model predicts accurate and unbiased output.
As a result, our network consists of tangent hyperbolic activation function and average pooling layers as depicted in \autoref{fig:network}.
We will elaborate the comparison of network configuration in \autoref{sec:ablation}.

\subsubsection{Configuration of input channels}\label{sec:input_config}
% input & output
Our data-driven solver takes all necessary fluid quantities as input.
This includes velocity derivatives to capture velocity differences with neighboring grid positions.
In addition, the conventional solver requires fluid volume for each cell and a phase indicator (solid or not) on the grid positions.
In summary, for the 2D grid, the input channels consist of four velocity derivative channels $\frac{\partial u}{\partial x}, \frac{\partial v}{\partial y}, \frac{\partial u}{\partial y}, \frac{\partial v}{\partial x}$ for the velocity field $\textbf{u}=(u,v)$, one channel for fluid volume, and one channel for the solid indicator.

\noindent\textbf{\textit{Velocity derivative channels.}} We use velocity derivatives instead of velocity itself in the construction of input channels because viscosity acts to reduce the velocity derivatives of the neighboring fluid. 
Moreover, using the velocity itself as input, the output would be varied by the velocity scale.
For example, during free fall, there should be no velocity change caused by viscosity, but the network prones to output varying values due to the accelerated velocity.
Hence, velocity derivative channels are employed for generalization across different velocity scales.

\noindent\textbf{\textit{Fluid volume channel.}} To convey the volume quantity to the CNN, we suggest using a fluid volume channel.
First, the signed distance from the fluid surface, constructed from a spherical surface of each particle, is computed for each corner (4 corners for 2D cell, and 8 for 3D cell) of the cell. From signed distances of two adjacent corner $d^{+}$ and $d^{-}$, edge occupancy of fluid is derived as $V_e = - (d^{-} / ( d^{+}-d^{-}))$. Then, the face (or volume) occupancy is approximated as the average of the occupancy of its 4 edges (or 6 faces in 3D).
For the generalization of the fluid volume, we normalized the volume by the factor of maximum volume of one grid cell, so that the fluid volume is in [0, 1].

\noindent\textbf{\textit{Solid indicator channel.}} In addition, to deliver the status of the grid position, we suggest solid indicator channel, constructed by the signed distance ${D}_{i,j}$ from solid on the position $\textbf{x}=(i,j)$.
We set the solid indicator channel as follows:
\begin{equation}\label{equ:solid_indicator}
\textrm{solid indicator}_{i,j} = \left\{
\begin{array}{ll}
0 \ & \textrm{if ${D}_{i,j} > 0$}\\
1 \ & \textrm{if ${D}_{i,j} \le 0$}\\
\end{array}
\right.
\end{equation}

\noindent\textbf{\textit{Viscosity coefficient channel.}} Finally, we suggest an optional viscosity coefficient channel.
While the network converges well without this channel when trained for a fixed viscosity coefficient, it becomes essential for generalization across a wide range of viscosity coefficients. The viscosity coefficient channel, denoted by coeff$_{i,j}$, is assigned the dynamic viscosity $\mu$ in Pa·s. When implementing the channel, only cells with positive fluid volume are set as $\mu$. The viscosity coefficient channel is set as follows:
\begin{equation}
\textrm{coeff}_{i,j} = \left\{
\begin{array}{ll}
0 & \textrm{if cell$_{i,j}$ is air or solid}\\ 
\mu & \textrm{if cell$_{i,j}$ is liquid}
\end{array}
\right.
\end{equation}

The U-shaped architecture requires pooling, thus we expand the dimension by padding to be divisible by $2^{n}$ where $n$ is the number of pooling layers.
Zero padding is applied for velocity derivative, fluid volume, and viscosity coefficient channels, while one is used for a solid indicator channel.
Random padding is applied to both the height and width dimensions to accommodate the various positions of the rigid body.

Our architecture readily extends to 3D fluid simulation using the symmetric MAC grid for 3D fluid quantities as shown in \autoref{fig:3d}. 
All layers of U-shaped architecture should be replaced into 3D layers, for example, 3D convolution, 3D average pooling, and 3D transposed convolution layers for upsampling.
The input channels are given with $\frac{\partial u}{\partial x}, \frac{\partial v}{\partial y}, \frac{\partial w}{\partial z},\frac{\partial u}{\partial y}, \frac{\partial u}{\partial z}, \frac{\partial v}{\partial x}, \frac{\partial v}{\partial z},\frac{\partial w}{\partial x}, \frac{\partial w}{\partial y}$, fluid volume, and solid indicator on symmetric MAC grid for fluid velocity $\textbf{u}=(u,v,w)$.

\subsection{Symmetric MAC grid for CNN}\label{sec:sym}
Marker-and-cell (MAC) grid \cite{MAC} stores fluid velocity and mass quantities at the edge of the grid as illustrated in \autoref{fig:figures}(a) so that the dimensions of the quantities vary from axis to axis.
For example, for the grid size $(n, n)$, the velocity dimension of the x-direction is $(n+1, n)$ and the y-direction is $(n, n+1)$.
Therefore, to input information from the x and y-direction together, modifying the dimensions is required.
\begin{figure}[!h]
\centering
\includegraphics[width=0.6\linewidth]{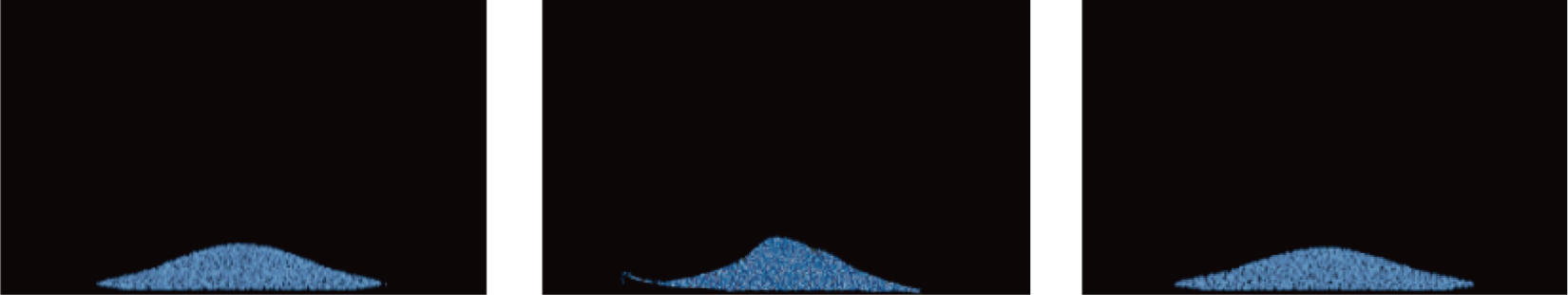}
\caption{\textbf{Comparison of input architecture.} (Left) Ground truth, (Middle) Result from the input with simple padding, and (Right) input with our symmetric MAC grid.}
\label{fig:tilt}
\end{figure}

For fluid simulation, fluid quantities must be handled with care since tiny biases accumulate errors over successive frames.
For the case of CNN for images, interpolation or padding could be a simple solution that rectifies the dimension.
However, interpolation yields a loss of fluid information.
Furthermore, simple padding causes an asymmetry between input channels that results in a tilted simulation, as \autoref{fig:tilt} illustrates.

To address this problem we suggest a symmetric MAC grid for CNN that preserves all information without loss and generates symmetric input channels, regardless of the axis.
Our symmetric MAC grid extends the original MAC grid including not only the centers but also the corners and edges of the cells.
Therefore, the dimension of the symmetric MAC grid becomes $(2n+1,\ 2n+1)$ for the $(n, n)$ grid.
The quantities of mass and velocity of the x-direction are defined in pink circles, and the y-direction in blue circles as shown in \autoref{fig:figures}(b).
The network output channels, which are the x-direction and the y-direction velocity change, are also placed in the same position.
The other cells without value are filled with zero not to influence the neural network.

In addition, this architecture facilitates the generation of derivative channels. 
To be more specific, $\frac{\partial u}{\partial x}, \frac{\partial v}{\partial y}$ are placed in the yellow cell and $\frac{\partial u}{\partial y}, \frac{\partial v}{\partial x}$ are in the gray cell.
Liquid volume and solid indicator channels are calculated for all cells.
Therefore, our input architecture guarantees that all input channels are symmetric and yields symmetric simulation results as shown in \autoref{fig:tilt}.

\section{Experiments on U-shaped CNN architecture}\label{sec:ablation}
\begin{figure}[h]
    \centering
    \includegraphics[width=0.45\textwidth]{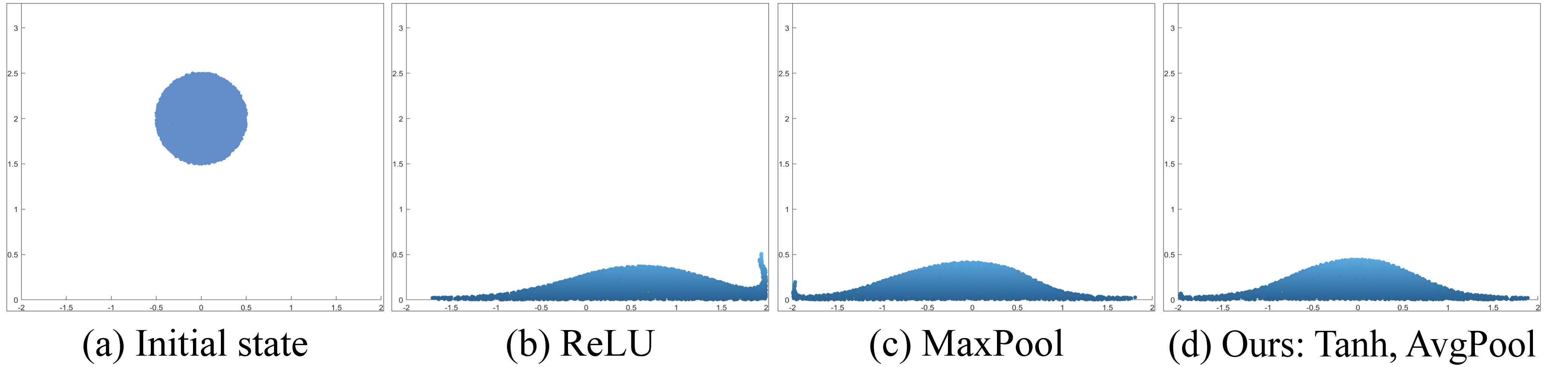}
    \caption{(a) From the initial resting state, the frame with (b) average pooling and ReLU (c) max pooling and tangent hyperbolic (Tanh) (d) average pooling and Tanh (ours).}
    \label{fig:abl_layer}
\end{figure}
\begin{figure}[h]
    \centering
    \includegraphics[width=0.45\textwidth]{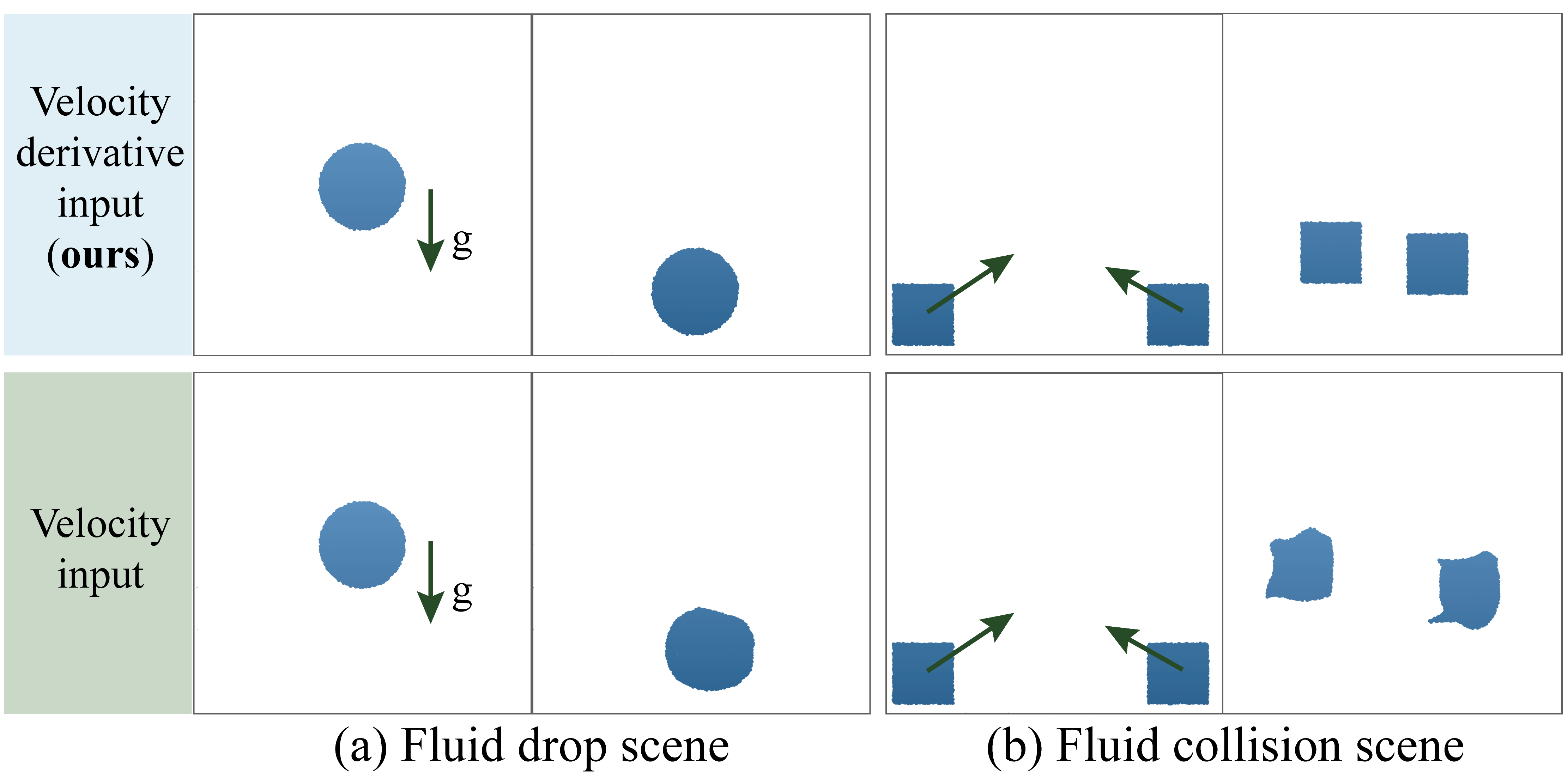}
    \caption{\textit{Top row:} velocity derivative input and \textit{Bottom row:}  velocity input for (a) fluid drop and (b) colliding fluids scenes. The scenes are in a training dataset.}
    \label{fig:vel}
\end{figure}

Ablation studies on network architecture are demonstrated in this section.
We verify our network for various conditions including activation functions, pooling methods, and input channels.

\noindent\textbf{\textit{Activation function and pooling layers}}
The most important factor in network configuration is unbiased learning because bias results in tilted simulation.
To explore the best CNN layers, we conducted ablation studies for the activation function between ReLU and tangent hyperbolic, and the pooling method between the max and average pooling.
As shown in \autoref{fig:abl_layer}, ReLU which removes negative features, and max pooling layer which ignores small features produce a distorted result for fluid drop scene.
Therefore, we design our network with a tangent hyperbolic activation function and average pooling layers.

\begin{figure*}[t]
\centering
\includegraphics[width=0.85\linewidth]{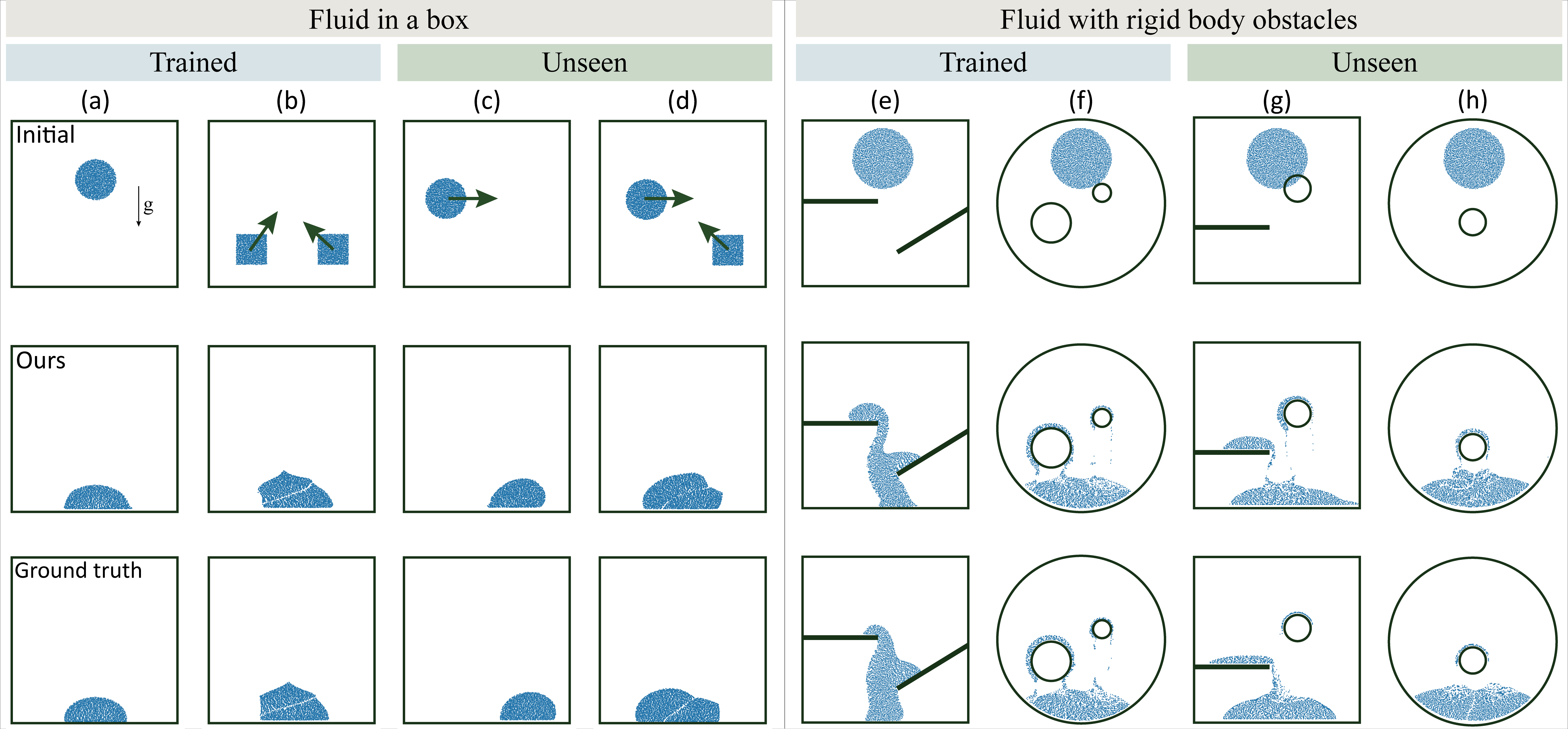}
\caption{Simulation result comparison between our data-driven solver and ground truth for various scenarios}
\label{fig:res}
\end{figure*}

\begin{figure}[t]
\centering
\includegraphics[width=0.75\linewidth]{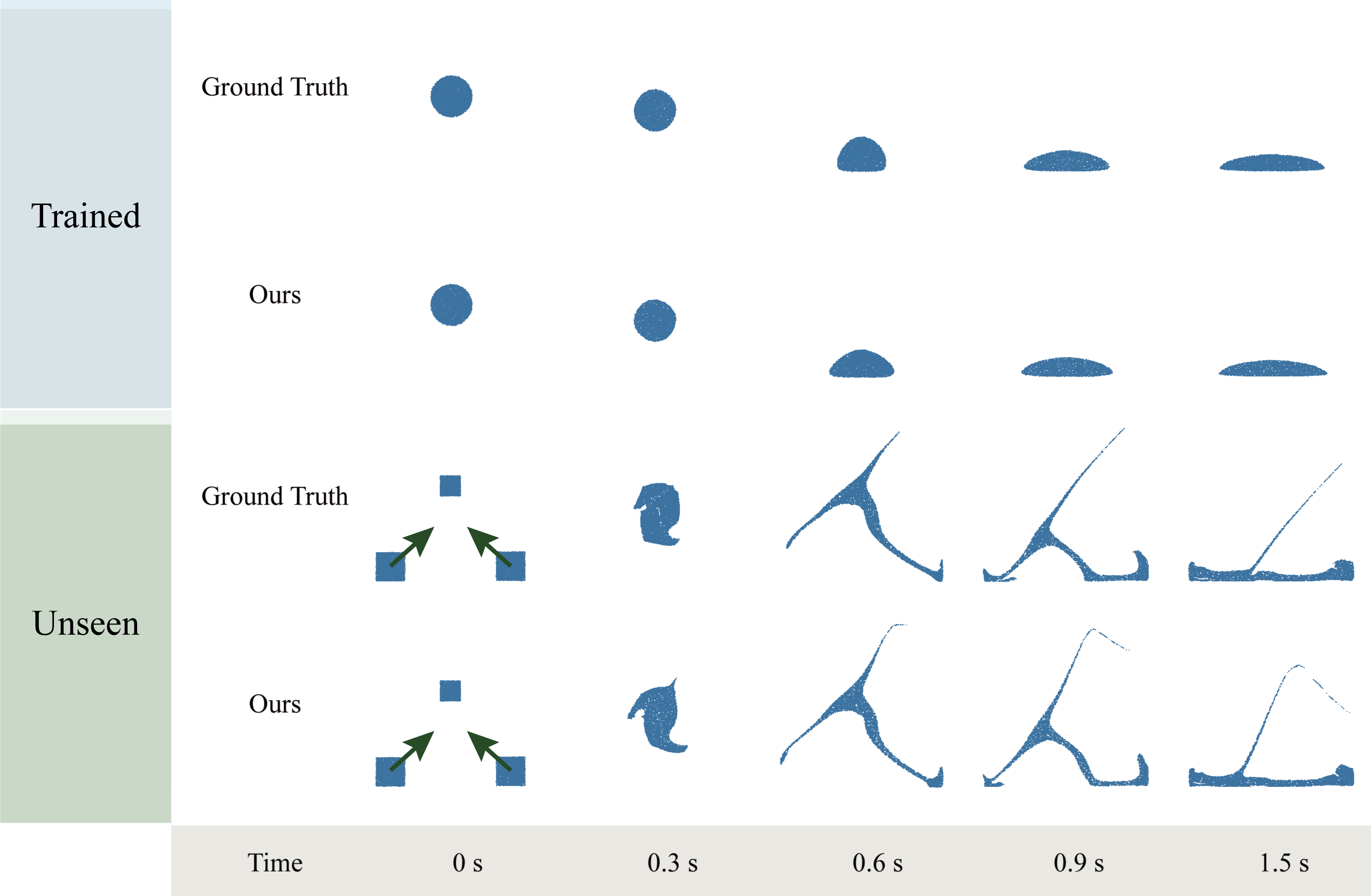}
\caption{Simulation result comparison with timestamps for both trained and unseen scenes}
\label{fig:res2}
\end{figure}

\noindent\textbf{\textit{Velocity channel vs. velocity derivative channel}}\label{sec:ablation_vel}
As mentioned in \autoref{sec:input_config}, the network receives velocity derivatives instead of raw velocity inputs.
As shown in \autoref{fig:vel}(a), the velocity derivative maintains the shape of the sphere during free fall, whereas the velocity channel distorts the sphere. 
This is because the velocity is accelerated, so the CNN solver prones to produce different outputs across the time steps.
On the other hand, since the velocity derivative has a constant value of zero during free fall, the consistent output is generated.
For the same reason, the velocity channel also produces a strange behavior for a more dynamic scene such as \autoref{fig:vel}(b).

\begin{figure}[t]
\centering
\includegraphics[width=0.6\linewidth]{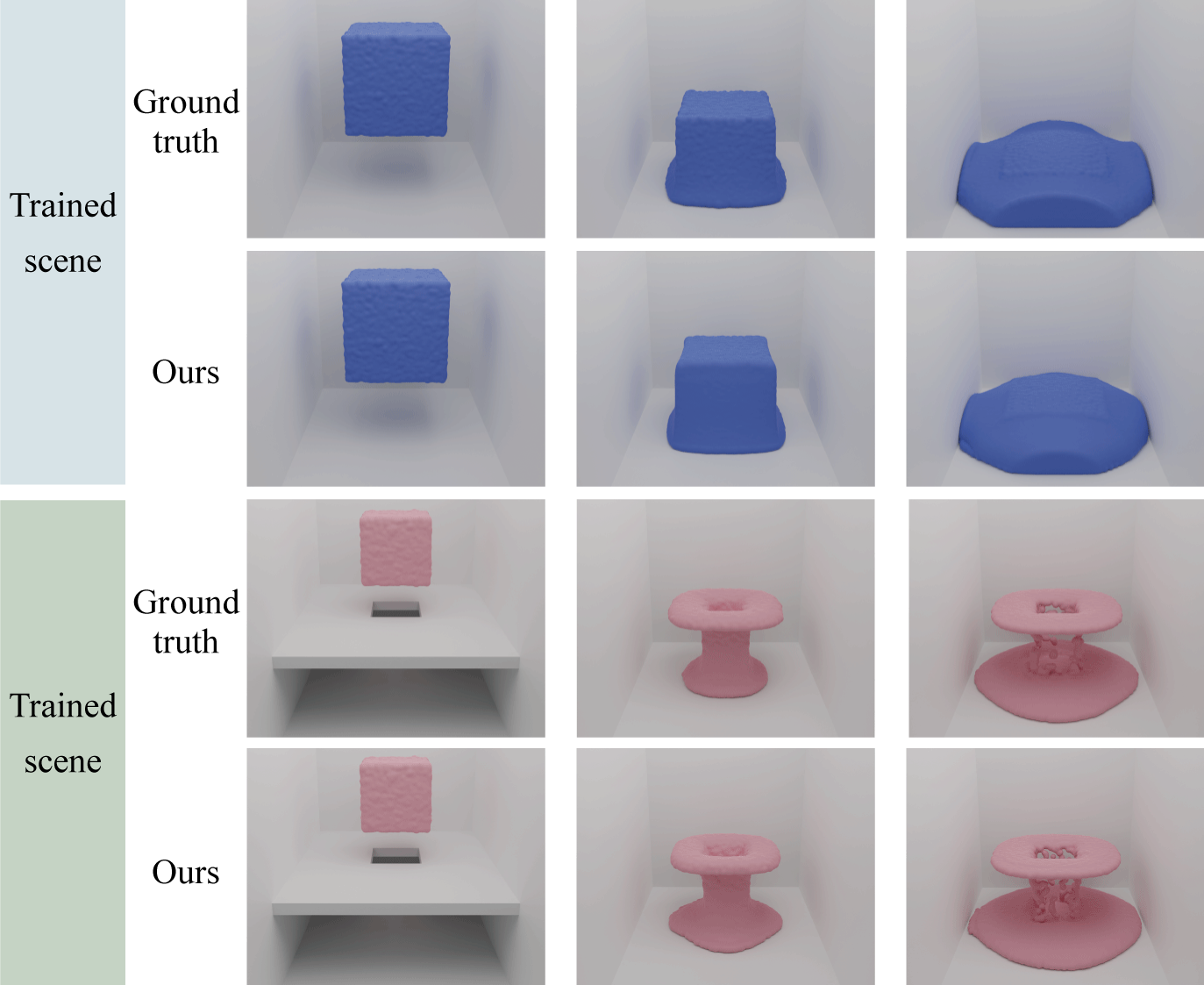}
\caption{\textbf{3D fluid-rigid simulation} Trained solely on the fluid drop scene, our solver similarly simulates the unseen scene with the ground truth. The space is $2m\times2m\times2m$ with $80\times80\times80$ Eulerian grid dimension .}
\label{fig:3dresult}
\end{figure}

\begin{figure}[t]
\centering
\includegraphics[width=0.6\linewidth]{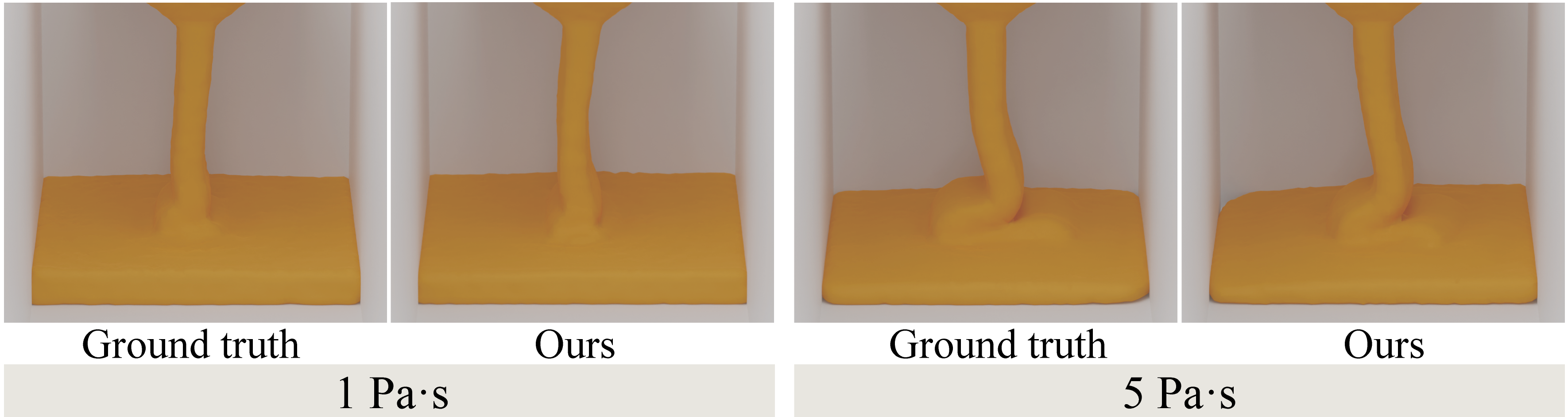}
\caption{Buckling effect of viscous fluid with 1 and 5 Pa·s dynamic viscosities. The simulated space is $0.6m\times1m\times0.6m$ with Eulerian grid dimension $48\times80\times48$.}
\label{fig:buckling}
\end{figure}
%%%%%%%%%%%%%%%%%%%%%%%%%%%%%%%RESULT&DISCUSSION%%%%%%%%%%%%%%%%%%%%%%%%%%%%%%%
\section{Results}\label{sec:result}

\subsection{Universality of data-driven viscosity solver}
In order to verify the universality of the proposed architecture, we demonstrate fluid simulation with various scenarios: fluids in a box, fluids with rigid body obstacles, and 3D scenes exhibiting buckling effect.

\subsubsection{Viscous fluid with rigid bodies}
We demonstrate our solver with two scenarios: fluid in a box and fluid with rigid body obstacles in \autoref{fig:res}. Both scenes are constructed with 1 Pa·s dynamic viscosity, 300 Hz framerate, $2m\times2m$ space, and  $100\times100$ Eulerian grid dimension. The training epoch is 2000 and the learning rate is $5\times10^{-4}$. Our data-driven viscosity solver (with four pooling layers) takes approximately 0.007 seconds per frame. 

For the fluid in a box scenario, the neural network is trained using 5-second fluid simulations of \autoref{fig:res}(a,b), resulting in a dataset consisting of 1500 frames for each scene. Our solver outputs acceptable visual fidelity for both trained and unseen(\autoref{fig:res}(c,d)) scenarios.

We tested our method in more complex scenes with rigid body obstacles.
The network is trained with 10-seconds fluid simulations of \autoref{fig:res}(e,f).
Our solver also shows similar results to ground truth for unseen scenes of \autoref{fig:res}(g,h).

For further validation of our data-driven simulation, we performed a timestamp-based comparison between ours and the ground truth. For both trained and unseen scenarios, our data-driven solver consistently predicts similar velocity changes for every timestamp as visualized in \autoref{fig:res2}.

\subsubsection{Extension to 3D fluid simulation}\label{sec:3d}

As depicted in \autoref{fig:3dresult}, our solver trained on a fluid drop scene successfully performs 3D fluid simulations for both trained and unseen scenes.
The network effectively learns viscosity characteristics using data from a simple scene and extends its capability in more complex scenes with rigid bodies.

Our viscosity solver (with two pooling layers) takes approximately 0.39 seconds per frame with an $80 \times 80 \times 80$ Eulerian grid for scenes in \autoref{fig:3dresult}.

Moreover, our solver simulates the buckling effect of viscous fluid, demonstrated in \autoref{fig:buckling}. Similar to the ground truth, the fluid with a lower viscosity coefficient (1 Pa·s) exhibits reduced buckling compared to the 5 Pa·s case.

\subsection{Generalization of dynamic viscosity}
\begin{figure}[h]
    \centering
    \includegraphics[width=0.6\linewidth]{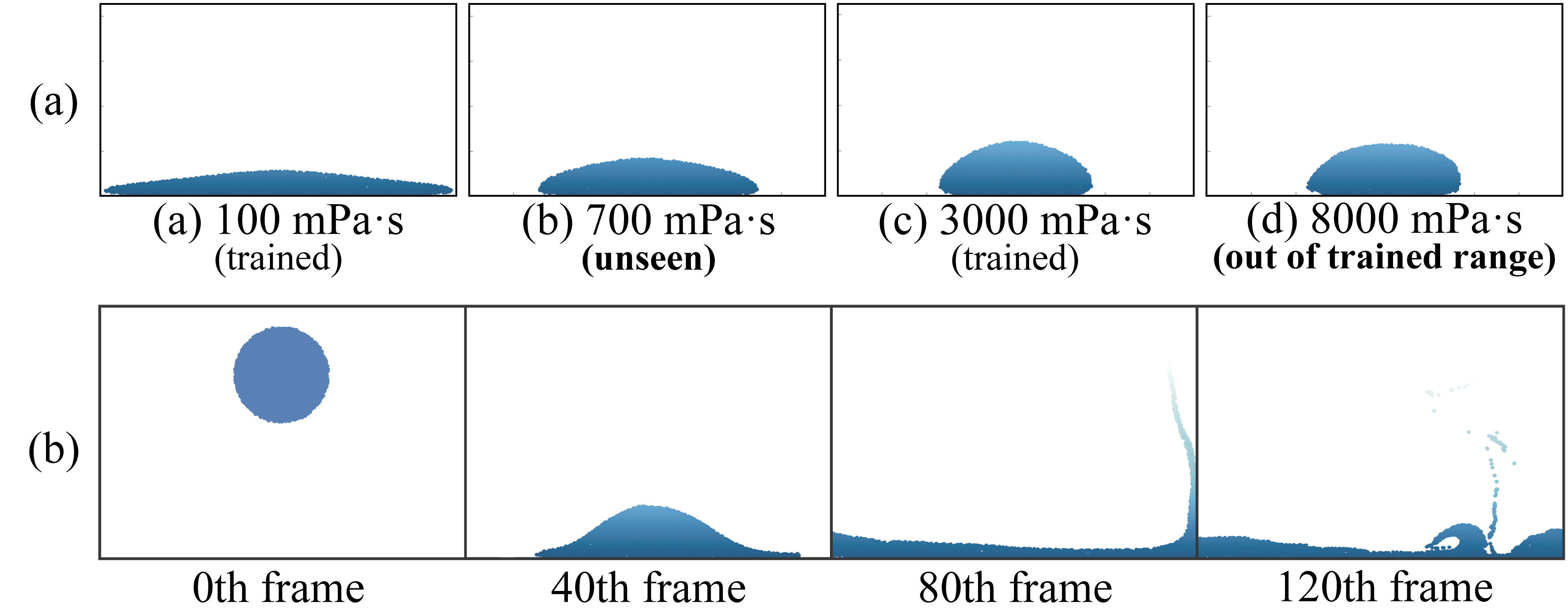}
    \caption{(a) Fluid drop scenes with one neural network model for various viscosity coefficients.(b) Simulation result with viscosity coefficient channel set by left side as 1000 and right side as 1 mPa·s.}
    \label{fig:coeff}
\end{figure}

Our network can be generalized to various dynamic viscosities $\mu$.
To provide dynamic viscosity information, the viscosity coefficient input channel is optionally given to the network as mentioned in \autoref{sec:input_config}.
U-shaped viscosity solver with four pooling layers is trained by 12000 frames consisting of dynamic viscosity 1, 100, 500, 1000, 2000, and 3000 mPa·s equally.
The trained model is well generalized for unseen coefficients as shown in \autoref{fig:coeff}(a). For example, although 700 mPa·s is not learned, it is spread narrower than 100 and wider than 3000.

However, as shown in \autoref{fig:coeff}, our data-driven solver cannot extrapolate the viscosity simulations outside the trained dynamic viscosity range.
The simulation is tilted to the right and upside because a large dynamic viscosity affects the network to output large x-direction (right) and y-direction (upward) velocity changes.
Therefore, a corresponding dataset is required to generate the model for a wide range of dynamic viscosity.

Moreover, mixed dynamic viscosity can be simulated without additional training. For example, we assigned dynamic viscosity 1000 to the left and 1 mPa·s to the right side in \autoref{fig:coeff}(b). The left side is less spread out due to the large viscosity, whereas the right side fluid spreads like water.
We anticipate that our model can be applied to complex scenes with multiple viscosity coefficients. In-depth exploration of multiple viscosity coefficients remains a future work.

\subsection{Real-time application of rigid-fluid interaction}\label{sec:real-time}

\begin{figure}[h]
    \centering
    \includegraphics[width=0.6\linewidth]{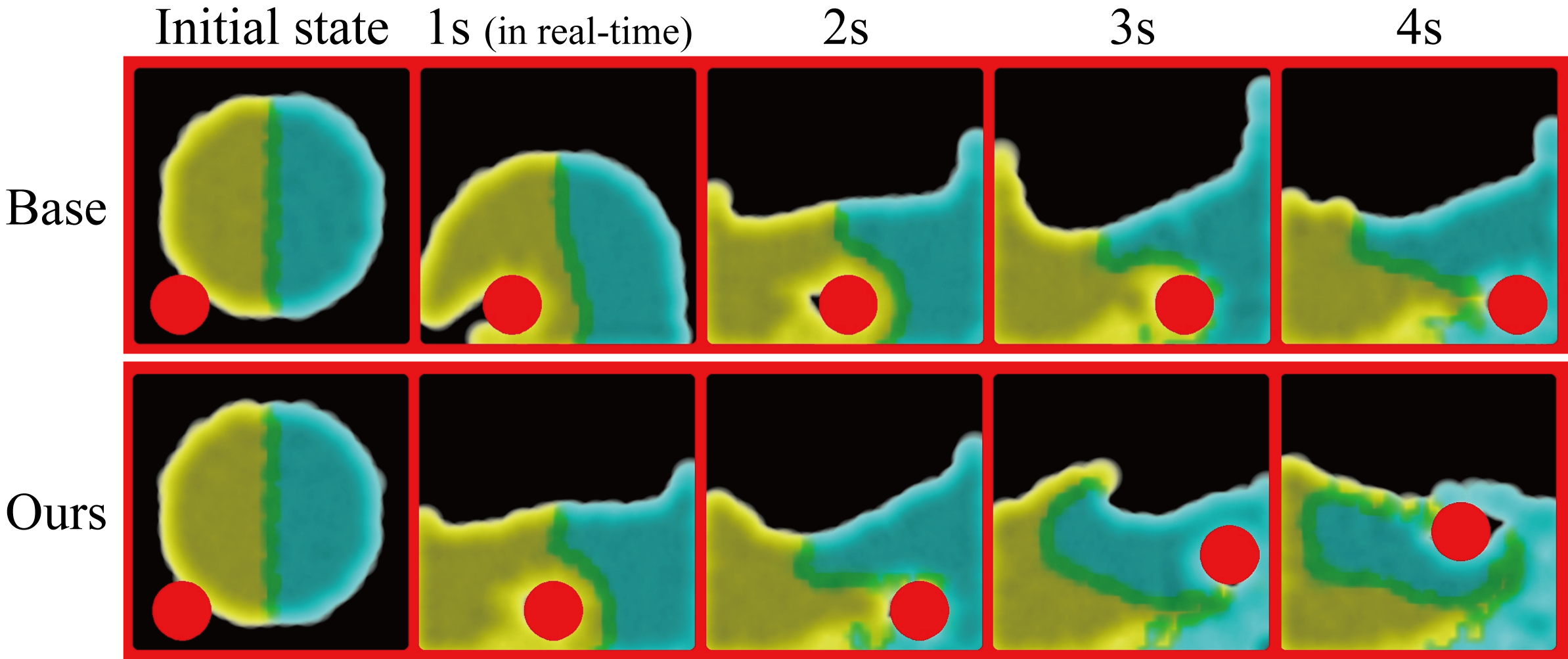}
    \caption{Progression of real-time 2D rigid-fluid interactive application. \textit{Top row: } Iterative viscosity solver \cite{Batty}. \textit{Bottom row: } Our U-shaped CNN viscosity solver which is nearly twice faster.}
    \label{fig:rtapp}
\end{figure}
To demonstrate the efficiency of our solver, we applied our solver to a real-time interactive application of paint mixing \autoref{fig:rtapp} where a rigid pointer (red circle) perturbs fluid of two different colors.
The scene consists of $(25, 25)$ grid dimension and $848$ particles.
On average, the baseline \cite{Batty} runs at 14 Hz (71 ms) with 32ms iterative viscosity solver, and ours takes 23Hz (42 ms) with 4ms data-driven viscosity solver.

%%%%%%%%%%%%%%%%%%%%%%%%%%%%%%%CONCLUSION%%%%%%%%%%%%%%%%%%%%%%%%%%%%%%%

\section{Conclusion}
In this paper, we propose a data-driven viscosity solver for a hybrid Lagrangian/Eulerian fluid simulator, utilizing a symmetric MAC grid to represent fluid quantities without symmetry breaking or interpolation. Our U-shaped CNN, with input channels of velocity derivatives, fluid volume, and solid indicators, predicts velocity changes with reasonable error in both trained and unseen scenes. To show the effectiveness of our architecture, we demonstrate various scenes, including fluid with rigid bodies, mixed dynamic viscosity, and 3D fluid with a buckling effect.

We believe that the proposed data-driven viscosity solver can be utilized for diverse fluid simulation, as demonstrated in a real-time interactive application. Our solver's generalizability across a wide viscosity range is achieved through a viscosity coefficient channel and a physics-based variational loss.
While we primarily analyze our network in 2D space, we plan to further explore and optimize our method for 3D viscous fluid simulation.

\section*{\uppercase{Acknowledgements}}

This work was supported by Institute for Information \& communications Technology Promotion(IITP) grant funded by the Korea government(MSIT) (No.00223446, Development of object-oriented synthetic data generation and evaluation methods).

\bibliographystyle{alpha}
\bibliography{sample}

\end{document}